\documentclass[pra,twocolumn,showpacs,preprintnumbers,amsmath,amssymb]{revtex4}

\usepackage{graphicx}

\usepackage{dcolumn}
\usepackage{bm}
\usepackage{mathrsfs}
\newcommand{\eq}[1]{Eq.~(\ref{#1})}
\newcommand{\fig}[1]{Fig.~\ref{#1}}
\newcommand{\be}[1]{\begin{equation}\label{#1}}
\newcommand{\ee}{\end{equation}}
\renewcommand{\vec}{\mathbf}
\def\1E{\mathrm{I}}
\def\2E{\mathrm{II}}

\begin{document}
\title{Attosecond time-scale multi-electron col
lisions in the Coulomb four-body problem: traces in classical probability densities}
\author{Agapi Emmanouilidou$^{1,2}$ and Jan Michael Rost$^{1,3}$}
\address{$^{1}$ KITP, University of California Santa Barbara, Santa Barbara, CA 93106\\
$^{2}$ ITS, University of Oregon, Eugene, Oregon 97403-5203\\
$^{3}$Max Planck Institute for the Physics of Complex Systems, 01187 Dresden, Germany }
\date{\today}

\begin{abstract}In the triple ionization of the Li ground state by
single photon absorption the three electrons escape to the continuum
mainly through two collision sequences with individual collisions
separated by time intervals on the attosecond scale.  We investigate the
traces of these two collision sequences in the classical probability
densities.  We show that each collision sequence has characteristic
phase space properties which distinguish it from the other.  Classical
probability densities are the closest analog to quantum mechanical
densities allowing our results to be directly compared to quantum
mechanical results.
\end{abstract}
\maketitle
 
\section{Introduction}

The theoretical treatment of multiple ionization processes by single
photon absorption is highly complex with no analytic solution.  In the
energy domain, the difficulty is that one has to account for the
correlated motion of the electrons in the asymptotic form of the final
continuum state.  In the time domain, this difficulty can be avoided
at the expense of propagating the fully coupled few-body Coulomb
problem in time.

 One can surmount the obstacles in the
 theoretical treatment of the triple photo-ionization from the ground
 state of Lithium, for a wide range of energies, by formulating the
 four-body break-up process quasiclassically \cite{ER1}.  This implies
 classical propagation of the Coulomb four-body problem using the
 classical trajectory Monte Carlo (CTMC) phase space method.  CTMC has
 often been used to describe break-up processes induced by particle
 impact \cite{CTMC1, CTMC2, CTMC3, CTMC4} with implementations
 differing usually in the way the phase space distribution of the
 initial state is constructed.  We use a Wigner transform of the
 initial quantum wave function for the initial state, and this is why
 we call our approach ``quasi''-classical.  Naturally, the
 electron-electron interaction is treated to all orders in the
 propagation, and any difficulties with electron correlation in the
 final state are absent, since the method is explicitly
 time-dependent.  The results from the quasiclassical formulation for
 a wide range of energies \cite{ER1} were found to be in very good
 agreement with experimental results \cite{Wehlitz1,Wehlitz2} as well
 as theoretical ones available for higher excess energies
 \cite{Pindzola04,Pattard}.

Moreover, our classical results allow for a detailed analysis of the
physical processes in terms of the classical trajectories:  As we 
have demonstrated, the triply photo-ionizing trajectories can be organized in
groups according to the respective sequence of electron-electron
collisions \cite{ER2}.  According to this collision scheme we have
identified two main sequences that lead to triple ionization from the Li
ground state.  An indirect verification of the collision scheme could
be achieved by measurement of the electronic angular correlation
probability: we have shown for excess energies close to threshold
\cite{ER2}, that the classification scheme of ionizing trajectories
can explain the electronic angular correlation probability in terms of
the dominant ``T-shaped'' pattern of the three escaping electrons.  The
electronic angular correlation probability is not yet known
experimentally.  However, it should be measurable with state of the
art experimental techniques. 

In the current paper, we explore the manifestations in classical
probability densities of the two main collision sequences the
three electrons follow to ionize from the ground state of Li.  While
our previous treatment of the collision sequences was on the level of
single trajectories \cite{ER1,ER2} we now treat them on the level of
ensemble averages.  Our motivation for doing so is that probability
densities are the closest classical analog to quantum mechanics.  In
quantum mechanics the probability density is defined directly through
the quantum mechanical wave-function.  In classical physics probability
densities can be easily computed allowing for a visualization of the
differences between classical and quantum mechanical observables and
for exploring the correspondence principle.  Our ideas should be a
useful tool for identifying and understanding collision mechanisms in
other systems where collision processes play an important role, e.g., 
strongly driven atomic systems.

 Finally, the collision sequences in triple ionization of Li take
 place on an {\it attosecond} time scale.  While the first collision,
 in each ionization path, occurs around a couple of attoseconds after 
 photoabsorption the
 second collision takes place around $70$ attoseconds.  This is
 another interesting aspect of our work: the emergence of attosecond
 laser pulses represents one of the most exciting developments in
 ultrafast laser science over the past few years \cite{Atto1,Atto2}.
 Attosecond pulses raise the prospect of studying electronic
 wave-packet motion on the time scales at which this motion occurs in
 nature, namely, the atomic unit of time (1 a.u.=24 attoseconds).
 These time scales show why attosecond pulses are new tools for
 exploring electronic processes at their natural time-scale and at
 dimensions shorter than even atomic-dimensions.  The advancement of
 ultra-short laser science and its so far success in exploring and
 controlling electronic motion \cite{Krausz} renders  a future
 direct experimental verification of our collision sequences possible.

\section{Time-dependent quasiclassical description of ionization}
Typically, there are two structurally different contributions in
quantum mechanical matrix elements $\left <\psi_{f}|{\cal
O}(t)|\psi_{i}\right >$, the wavefunctions $\psi_{i,f}$ and the
operator ${\cal O}(t)$.  We call our description quasiclassical
because we use the full wavefunctions $\psi_{i}$ -- exactly translated
to a phase space density $\rho(\gamma)$ through a Wigner transform -- while  
the subsequent propagation of the density in time is performed fully 
classically.  Here, we are
interested in final states with all three electrons of Li in the
continuum.  Since we propagate the entire four body system over very
long times, we can simply project onto momentum states (corresponding
to a measurement at the detector).  In practise, this is done by
binning final momenta of trajectories, very similarly as in the
experiment.

\subsection{The initial phase space density for single photon multiple ionization}
The construction  of the initial 
 phase space density
$\rho(\gamma)$ in our quasiclassical formulation of the
triple photoionization of Li has been detailed in \cite{ER1}, here we 
give only a brief summary.  We
formulate the triple photoionization process from the Li ground state
($1s^{2}2s$) as a two step process \cite{Samson,Pattard,scch+02}.
First, one electron absorbs the photon (photo-electron) at time 
$t=t_{\rm abs}=0$.  Then, due to
the electronic correlations, redistribution of the energy takes place
resulting in three electrons escaping to the continuum.  It is the
latter step that we describe in our formulation.  We first assume that
the photo-electron is a $1s$-electron.  It absorbs the photon at the
nucleus ($\vec r_{1}=0$), an approximation that becomes exact in the
limit of high photon energy \cite{Kabir}.  The photon could also be
absorbed by the Li $2s$-electron.  However, the cross section for photon
absorption from a $1s$ orbital is much larger than from a $2s$ orbital
\cite{emsc+03}.  Hence, we can safely assume that the photo-electron
is a $1s$ electron which significantly reduces the initial phase space
to be sampled.  Also, by virtue of their different character the
electrons become practically distinguishable and allow us to neglect
antisymmetrization of the initial state.  We denote the photo-electron
by 1, the other $1s$ electron by 2 and the $2s$ electron by 3.
Immediately after photon absorption, we model the initial phase space
distribution of the remaining two electrons, $1s$ and $2s$, by the
Wigner transform of the corresponding initial wavefunction $\psi({\bf
r}_{1}=0,{\bf r}_{2},{\bf r}_{3})$, where ${\bf r}_{i}$ are the
electron vectors starting at the nucleus.  We approximate the initial
wavefunction as a simple product of hydrogenic orbitals
$\phi^{\mathrm{Z}_{i}}_{i}(\vec r_{i})$ with effective charges
$Z_{i}$, to facilitate the Wigner transformation.  The $Z_{i}$ are
chosen to reproduce the known ionization potentials $I_{i}$, namely
for the 2s electron $Z_{3}=1.259$ ($I_{3}=0.198\,$a.u.) and for the 1s
electron $Z_{2}=2.358$ ($I_{2}=2.780\,$a.u.).  (We use atomic units
throughout the paper if not stated otherwise.)  The excess energy,
$E$, is given by $E=E_{\omega}-I$ with $E_{\omega}$ the photon energy
and $I=7.478$ a.u.\ the Li triple ionization threshold energy.
Following these considerations, the initial phase space density
is given by
\begin{equation}
\label{eq:distribution}
\rho(\gamma) = \mathscr{N}
\delta(\vec{r}_1)\delta(\varepsilon_{1}+I_{1}-\omega)\prod_{i=2,3}W_{\phi^{\mathrm{Z}_{i}}_{i}}
(\vec r_{i},\vec p_{i})\delta(\varepsilon_{i}+I_{i})
\end{equation}
with normalization constant $\mathscr{N}$.
 
To determine which fraction of $\rho(\gamma)$ leads to triple 
ionization, the phase space distribution must be followed in time.


 \subsection{The evolution of classical phase space densities}
 
 The evolution of a classical phase space density is determined by 
 the classical Liouville equation, which may be obtained within the 
 quantum mechanical phase space Wigner formalism  \cite{Wignertran} by taking the 
  limit $\hbar=0$
 \cite{Heller, Geyer},  
 \begin{equation}
 \label{eq:Liouville}
 \frac{\partial \rho(\Gamma(t))}{\partial t}=\mathscr{L}_{\mathrm{cl}}\rho(\Gamma(t)).
\end{equation}
The initial phase space values are
\begin{equation}
 \label{eq:gamma}
\Gamma(0) \equiv \gamma\,,
\end{equation}
and $\mathscr{L}_{\mathrm{cl}}$ is the classical Liouville operator
which is defined by the Poisson bracket \{H, \}, with H the
Hamiltonian of the system.  In our case H is the full Coulomb
four-body Hamiltonian.  In practice, \eq{eq:Liouville} amounts to
discretizing the initial phase space, assigning weights to each
discrete point $\gamma_{j}=(p_{j}(0),q_{j}(0))$ according to
$\rho(\gamma_{j})$, and evolving in time each initial condition
$\gamma_{j}$ with the Coulomb four-body Hamiltonian.  This amounts to
propagating electron trajectories using the classical equations of
motion (CTMC).  Regularized coordinates \cite{regularized} are used to
avoid problems with electron trajectories starting at the nucleus.  

\subsection{Different triple ionizing collision sequences and their 
phase space ensembles}

An important finding of our previous studies \cite{ER2} is that the
triple ionizing trajectories can be organized in classes according to
their ionization-driven properties.  In particular, we found two main
classes consisting of those trajectories that triply ionize through
the (12,13) collision sequence and those that ionize through
the (12,23) collision sequence.  In the first class the path
to ionization proceeds with photo electron 1 knocking out,
successively, electrons 2 and 3.  In the second class the
photo-electron 1 first knocks out electron 2 and then, electron 2
knocks out electron 3. More abstractly speaking, each class defines
an ensemble of trajectories which we label $\alpha = \1E$ and $\alpha = 
\2E$ for the (12,13) and the (12,23) collision sequences, respectively.

 For completeness, we briefly describe what we
 define as a momentum transferring electron-electron collision along a
 trajectory with time dependent electron positions $\vec
 r_{i}(t),\,i=1,2,3$ (see \cite{ER2}).  The term responsible for
 momentum transfer between electrons $i$ and $j$ is their Coulomb
 repulsion $V(r_{ij})=r^{-1}_{ij}$, $\vec r_{ij}=\vec r_{i}-\vec
 r_{j}$.  Hence, we identify a collision between electron $i$ and $j$
 ($ij$) through the momentum transfer 
 \be{mom} \vec D_{ij}: =
 -\int_{t_{1}}^{t_{2}}\nabla V(r_{ij})\,dt  \equiv \int_{t_{1}}^{t_{2}} {\vec F}_{ij} \,dt \ee 
 under the condition
 that V$(r_{ij}(t_{k}))$, $k = 1,2$ are local minima in time with
 $t_{2}>t_{1}$, while $r_{ij}=|\vec r_{i}-\vec r_{j}|$.  This
 automatically ensures that the integral of \eq{mom} includes the
 ``collision'' with a local maximum of V$(r_{ij}(t))$ at a time
 $t_{1}<t_{M}<t_{2}$.  During the time interval $t_{1}<t<t_{2}$, all
 four particles interact with each other.  Hence, the definition
 \eq{mom} is only meaningful if the collision redistributes energy
 dominantly within the subsystem given by the three-body
 Li$^{+}$-Hamiltonian, $H_{ij}$, of the nucleus and the electrons $i$
 and $j$ involved in the actual collision,
 \begin{equation}
     \label{he-ham1}
 H_{ij}=H_{i}+H_{j}
 +\frac{1}{\left |\vec r_{i}-\vec r_{j}\right 
 |}\,,
 \end{equation}
 with 
\begin{eqnarray}
 \label{he-ham2}
\dot H_{ij} \equiv \frac{dH_{ij}}{dt}\approx 0 &
\mathrm{for}\,\, t_{1}<t<t_{2} \,,
\end{eqnarray}
  where
 \begin{equation}
     \label{eq:hyd}
 H_{i}=p_{i}^2/{2}-Z/r_{i}
 \end{equation}
 are hydrogenic two-body 
 Hamiltonians with charge $Z=3$ of the Lithium ion.

Our goal is to investigate whether the two main ionization sequences we
have previously identified using \eq{mom}, manifest themselves on an
ensemble average level with properties that clearly distinguish one
from the other, thus reinforcing the validity of our classification
scheme.  To this end we need classical observables defined over 
arbitrary phase space ensembles $\alpha$, in our case the two 
ensembles $\1E$ and $\2E$.

\subsection{Classical probability densities for observables over a 
classical phase space ensemble}

 The probability density
$\mathscr{P}_{\alpha}(a,t)$ to find the value $a$ for the
observable $A$ at time $t$ under the ensemble $\alpha$
is given by
\begin{equation}
\label{eq1:probdens}
\mathscr{P}_{\alpha}(a,t)=\int_{\alpha} 
\delta(a-A(\gamma,t))\rho(\gamma)
{\rm d}\gamma,
\end{equation}
where $\int_{\alpha}\rm d\gamma$ denotes integration over 
initial phase space which contains only those
trajectories that belong to the ensemble $\alpha$.  The propagation begins at the time
$t_\mathrm{abs}=0$ of photoabsorption.  \eq{eq1:probdens} amounts to a)
propagating all the trajectories of the ensemble $\alpha$  from time 
$t_{\rm abs}$ up to $t$, b) computing for each
trajectory the observable $A(t)$, c) selecting only those trajectories
which satisfy $A(t)=a$ and adding together their weights.  Note,
that the probability to find at time $t$ the value $a$ for the
observable $A(t)$ is given by $\mathscr{P}_{\alpha}(a,t){\rm d} a$.
 
Finally, the classical average of the observable $A(t)$ over the
ensemble $\alpha$ is simply \cite{Rost}
\begin{equation}
\label{eq3:classobserv}
\left<A(t)\right>_{\alpha}=\int_{\alpha} A(\gamma,t)\rho(\gamma){\rm 
d}\gamma.
\end{equation}

\section{Ensemble averages of energy and the main collision sequences} 
\begin{figure}
\includegraphics[width=\columnwidth]{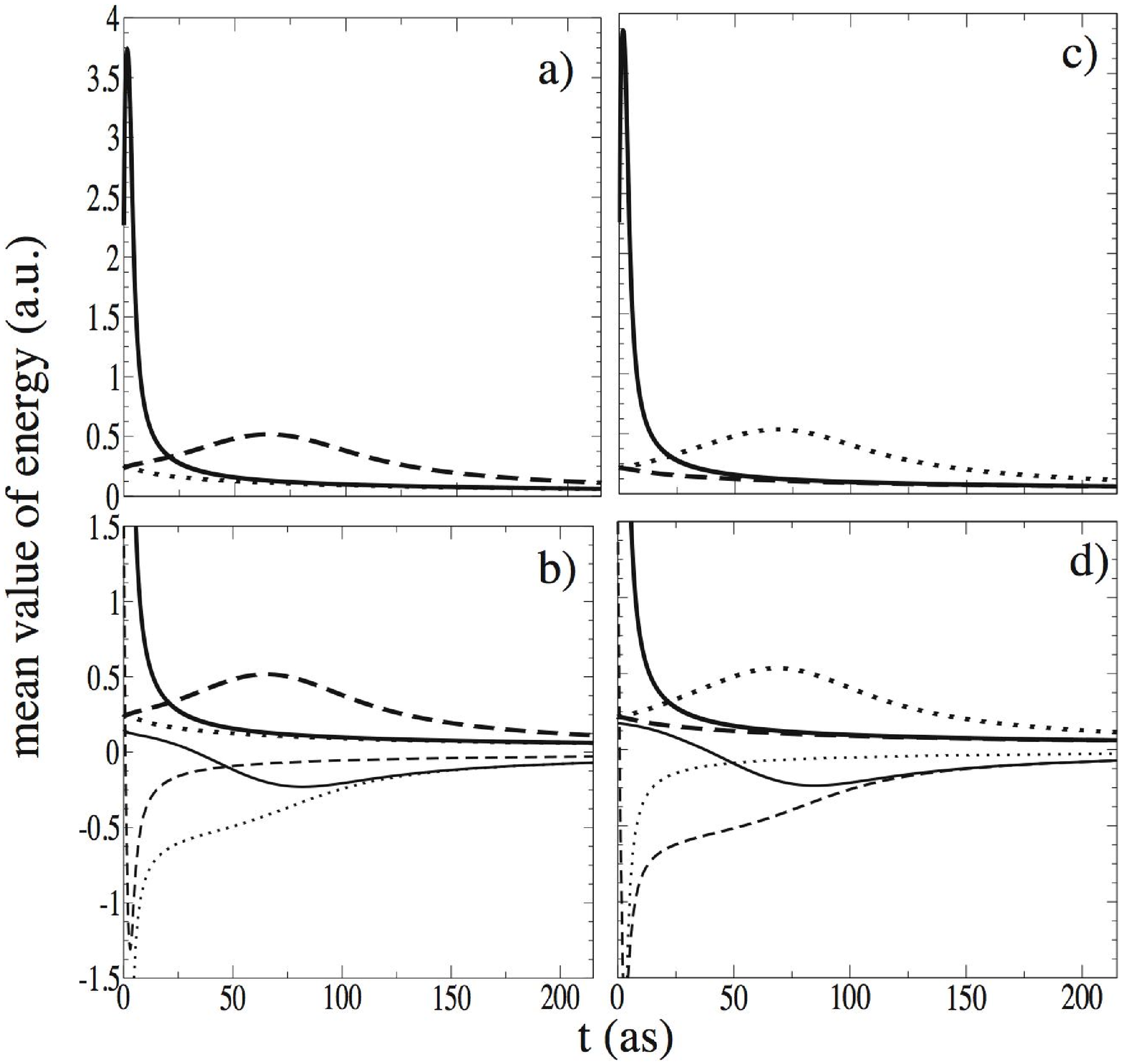}
\caption{\label{fig1:interEnergy} 
Averages over ensembles   $\alpha=\1E$ (left, a)--b))
and $\alpha=\2E$ (right, c)--d)), see \eq{eq3:classobserv}.
The upper panels show the
interelectronic repulsions,
$\left<1/r_{12}\right>_{\alpha}$ (solid),
$\left<1/r_{13}\right>_{\alpha}$ (dashed),
and $\left<1/r_{23}\right>_{\alpha}$ (dotted).
The lower panels show in addition the three-body energies
with thin lines
$\left<H_{12}\right>_{\alpha}$ (solid),
$\left<H_{13}\right>_{\alpha}$ (dashed),
$\left<H_{23}\right>_{\alpha}$ (dotted).
}
 \end{figure}

 That the ensembles $\1E$ and $\2E$, defined by the two main collision
 sequences, leave different traces on classical averages is obvious
 from \fig{fig1:interEnergy}.  This and all other results have been
 obtained at an excess energy of $E = 0.9\,$eV, that is fairly close to
 threshold \cite{ER1}. 
 Firstly, we see in \fig{fig1:interEnergy}a
 that the electron pair potential energies $\left
 <1/r_{12}\right>_{\1E}$ (solid line) and $\left
 <1/r_{13}\right>_{\1E}$ (dashed line) have well defined maxima at
 $t_{12}=1.6$ and $t_{13}=65$ attoseconds,
 even when averaged over all
 trajectories that ionize through the $(12,13)$ collision sequence.
 Note that $\left <1/r_{23}\right>_{\1E}$ (dotted line) decreases
 monotonically.  Thus, all trajectories triply ionizing through the
 $(12,13)$ collision sequence, satisfy as an ensemble the first
 criterion of this sequence, namely, maxima in the potential energies
 of the electron pairs 12 and 13 participating in collisions.  They
 also satisfy as an ensemble the second criterion, that is, while the
 $\left<1/r_{12}\right>_{\1E}$ potential energy changes during the 12
 collision the energy of the three-body Hamiltonian
 $\left<H_{12}\right>_{\1E}$, see Eqs.~\ref{he-ham1}, \ref{he-ham2},
 remains constant and while the $\left<1/r_{13}\right>_{\1E}$
 potential energy changes during the 13 collision the energy of the
 three-body Hamiltonian $\left<H_{13}\right>_{\1E}$ remains constant.
 This is clearly demonstrated in \fig{fig1:interEnergy}b:
 $\left<H_{12}\right >_{\1E}$ (thin solid line)
 remains almost constant near $t_{12}= 1.6$ as,
 while it changes around $t_{13}= 65 $ as, as should be the case since
 during the 13 collision it is the $\left<H_{13}\right> _{\1E}$ energy
 (thin dashed line)
 that is conserved.  Similarly, $\left<H_{13}\right >_{\1E}$ remains
 constant around $t_{13}=65$ as, while it changes near $t_{12}=1.6$ as,
 since during the 12 collision it is the $\left<H_{12}\right >_{\1E}$
 energy that is conserved.  We also plot $\left<H_{23}\right >_{\1E}$
 (thin dotted line)
 to demonstrate that it changes both, around the $t_{12}$ and $t_{13}$
 collision times.  One may summarize the two criteria for triple 
 ionizing ensembles of trajectories as
 \begin{itemize}
     \item[(A)]
     A maximum in time of the ensemble average $\langle 
     1/r_{ij}\rangle_{\alpha}$ defines a collision and its 
     time $t_{ij}$ between electrons $i$ and $j$. 
     \item[(B)]
     Near the time of collision $t_{ij}$ the corresponding three-body 
     energy $\langle H_{ij}\rangle_{\alpha}$ of the ensemble remains 
     approximately constant.
\end{itemize}     
 These criteria apply also  to ensemble $\2E$ for the collision 
 sequence (12,23) as one can see in
 Figs.~\ref{fig1:interEnergy}c and \ref{fig1:interEnergy}d. The
 collisions 12 and 23 are well defined and take place with maxima in
 $\langle 1/r_{12}\rangle_{\2E}$ at $t_{12}=1.7$ as  and
 in $\langle 1/r_{23}\rangle_{\2E}$ $t_{23}=69$ as, the corresponding 
 three-body energies remain almost constant around these times.
 
It is important to keep in mind that the collisions described above
are not binary but three-body collisions involving two electrons and
the nucleus.  If we were dealing with binary collisions then $\left
<p_{i}^2/2+p_{j}^{2}/2+1/\left |\vec r_{i}-\vec r_{j}\right | \right>$
instead of $\left<H_{ij}\right>$ would be constant during the $ij$
collision, which is not the case as one can easily show.  For
simplicity we will identify in the following a collision by $ij$.
However, it is always understood that the nucleus is part of the
collision and that $ij$ refers to the three-body Hamiltonian H$_{ij}$
as given in \eq{he-ham1}.
 
\section{Probability densities}
 \subsection{Position} 
\begin{figure}
\scalebox{0.35}{\includegraphics{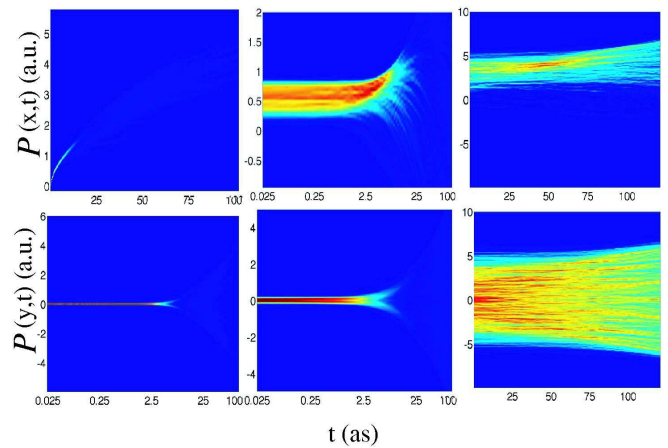}}
\caption{\label{fig2:x1wigner} Top panel: $\mathscr{P}_{\alpha}(x_{i},t)$ for
electrons $i=1,2,3$ (from left to right) for the $\alpha=\1E$ 
ensemble; Bottom panel: as for the top panel but for $\mathscr{P}_{\alpha}(y_{i},t)$.
 }
\end{figure}

  The probability densities for the Cartesian positions of each of the
  three electrons $i$, given by \eq{eq1:probdens} where $A(\gamma,t) =
  q_{i}(\gamma,t)$ with $q=x,y$ or $z$, do not offer much information
  regarding the identity of the electron pairs that participate in each collision, see
  \fig{fig2:x1wigner}.  However, they contain information about the
  electron pairs during the collisions: \fig{fig2:x1wigner} shows that
  the probability density of the Cartesian coordinates of electrons 2
  and 3 remain almost constant until $1.6$ and $65$ as, respectively,
  the times that electron 1 knocks them out in the 12 and 13
  collisions of the $(12,13)$ collision sequence.  In addition, the probability
  density of the $y$ and $z$ component of the photoelectron 1 does not
  change up to 1.6 as, the time of collision of electron 1 with 2.  This is expected for small times since the
  photoelectron's initial momentum is along the $x$-axis.  Similar
  conclusions can be drawn for the $(12,23)$ collision sequence.
 
 Due to our choice
 of initial conditions for the photoelectron (electron 1 always starts
 at $\vec r=0$ with initial momentum along the $x$-axis) our model has
 cylindrical symmetry around the $x$-axis.  As a result the probability
 density of $y_{i}$ is equal to that of $z_{i}$.  Moreover, since the
 invariance under the parity operation $y\to -y$ of the Hamiltonian is
 not broken through the initial conditions, \eq{eq:distribution}, the
 corresponding distributions $\mathscr{P}(y_{i},t)$ are symmetric
 about $y_{i}=0$, the same holds of course for the $z_{i}$-coordinates
 and for all times, see \fig{fig2:x1wigner}.  These symmetry
 properties extend to observables which respect them, such as, e.g.,
 individual electron momenta.

\subsection{Momentum}
The probability densities of the Cartesian momentum components for
each of the electrons allow for a deeper insight into the mechanism of
the electron collisions taking place during the two main sequences.
During an $ij$ collision the transfer of energy between electrons $i$
and $j$ is mediated through their mutual repulsion, $V(r_{ij})$.  As
discussed above, we have defined the collision time $t_{ij}$ as the
time $\dot{V}(r_{ij})=0$ with $V_{ij}$
maximal, see \eq{mom}.  However, a collision may formally be defined
to last between two minima of $V_{ij}$ at times $t_{1}<t_{ij}$ and
$t_{2}>t_{ij}$.  During the time the two electrons approach each
other, $t_{1}<t<t_{ij}$, $\dot{V}(r_{ij})>0$, while for
$t_{ij}<t<t_{2}$, $\dot{V}(r_{ij})<0$.

\begin{figure}
\scalebox{0.35}{\includegraphics{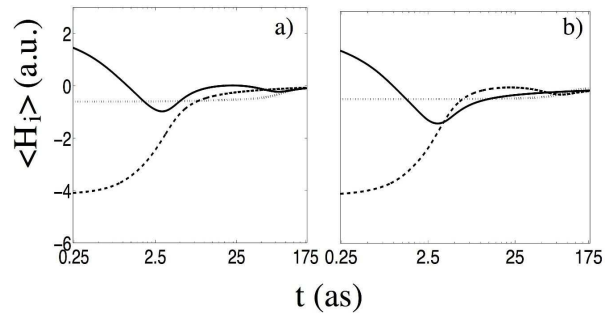}}
\caption{\label{wignerhammean} Single electron energy averages
$\left <H_{i} \right>_\alpha$ of electrons 1 (solid), 2 (dashed), 3 (dotted)
for the ensemble $\alpha =\1E$ ( a)) and $\alpha =\2E$
( b)).}
\end{figure}

This has different consequences for the corresponding time evolutions
of the individual hydrogenic energies $H_{i}(t)$ of the two electrons,
\eq{eq:hyd}, the one which suffers a net loss of energy in the
collision (the impacting electron) and the one which overall gains
energy in the collision (the impacted electron), see
\fig{wignerhammean}.  The latter gains energy throughout the
collision that is, $\dot{H}_{j}>0$ for
$t_{1}<t<t_{2}$.  On the other hand, the impacting electron looses
energy up to a time $t_{s}$, that is $\dot{H}_{i}<0$ for
$t_{1}<t<t_{s}$ with $t_{s}>t_{ij}$ ($t_{s}>t_{ij}$ follows from 
Eqs.~(\ref{he-ham1}, \ref{he-ham2}) and $\dot{V}(r_{ij})=0$ at time $t_{ij}$)
while for $t_{s}<t<t_{2}$ the impacting electron gains energy
$\dot{H}_{i}>0$.  From Eqs.~(\ref{he-ham1}) and (\ref{he-ham2}) we
can determine the change of a hydrogenic energy $\dot H_{j}$ in time 
during a collision with electron $i$. On the one hand we have
\be{dotHj}
\frac{dH_{j}}{dt} =
\vec p_{j} \cdot \dot {\vec p}_{j} +\frac{\partial H_{j}}{\partial\vec r_{j}} \cdot \vec p_{j}\,.
\ee
On the other hand we have 
 \be{dotHij}
  \dot {\vec{p}}_{j}\approx-\frac{\partial H_{ij}}{\partial
  \vec{r}_{j}}=\vec{F}_{ij}-\frac{\partial H_{j}}{\partial
  \vec{r}_{j}}\,.
  \ee
  Inserting \eq{dotHij} into \eq{dotHj} leads to
  \be{dotHj-2}
  \frac{d H_{j}}{dt}=\vec p_{j} \cdot \vec F_{ij}\,,
  \ee
  which shows, that the change of the hydrogenic energy of electron 
  $j$ 
  does not only depend on the modulus of the electron-electron force $\vec F_{ij}$ 
  but also on its direction relative to the momentum $\vec p_{j}$ of 
  electron $j$.
  We recall at this point that the present analysis of the
collision sequences in terms of the rate of change of the single
electron energies $H_{i}$ does by no means imply that we have 
calculated the evolution of trajectories with $H_{i}$.
All observables are evaluated with our numerical results
for the triple ionizing trajectories from the  propagated  full four-body 
Coulomb Hamiltonian, as we have already
pointed out in section IIA.

In \fig{wignerhammean} the ensemble averages $\langle
H_{i}\rangle_{\alpha}$ clearly illustrate the difference between the
impacting and the impacted electron for each three-body collision.
For trajectories from the ensemble $\alpha=\1E$ 
(\fig{wignerhammean}a)
electron 1 transfers
energy to electron 2 during the 12 collision as can be seen from the
sharp decrease of $\langle H_{1}\rangle_{\1E}$ followed by an increase
beginning at $t=2.4$ as while at the same time the energy $\langle
H_{2}\rangle_{\1E}$ of the impacted electron 2 increases.  The pattern
is repeated during the 13 collision where a decrease in $\langle
H_{1}\rangle_{\1E}$ is followed by an increase at $79$ as, while at
the same time  $\langle H_{3}\rangle_{\1E}$ of the impacted electron 3 is
increasing. 
The pattern of the hydrogenic energies during collisions is
also fullfilled for the ensemble $\2E$ as can be seen
in \fig{wignerhammean}b.

\begin{figure}
\scalebox{0.35}{\includegraphics{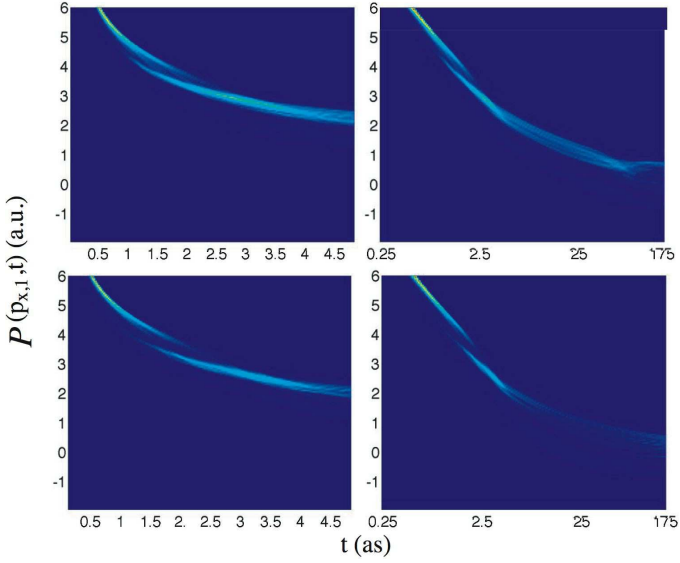}}
\caption{\label{fig7:px1} Momentum distributions $\mathscr{P}_{\alpha}(p_{x},t)$ for
electron 1  for the $\alpha =\1E$  (top) and $\alpha 
=\2E$ (bottom) ensemble. The left panels show the evolution for 
short times in greater detail.}
\end{figure}

\begin{figure}
\scalebox{0.35}{\includegraphics{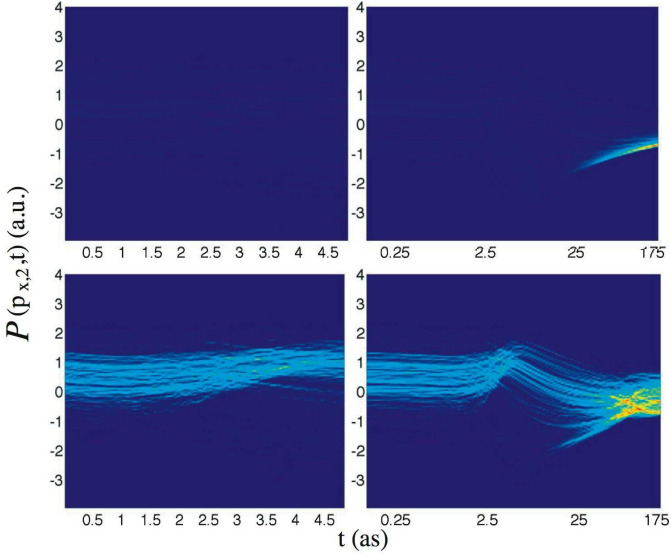}}
\caption{\label{fig9:px2} Same as Fig.~\ref{fig7:px1}
but for 
electron 2.}
\end{figure}

\begin{figure}
\scalebox{0.35}{\includegraphics{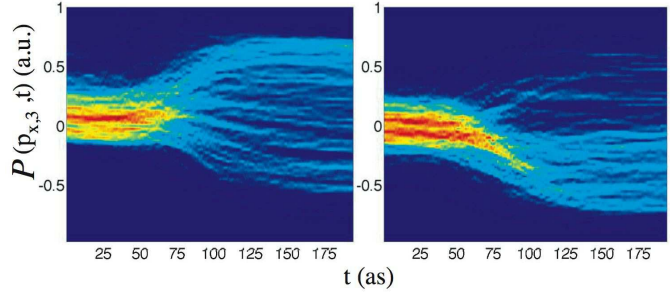}}
\caption{\label{fig11:pxy3}
Momentum distribution $\mathscr{P}_{\alpha}(p_{x},t)$
for electron 3 for the $\alpha =\1E$  (left) and $\alpha 
=\2E$ (right) ensemble.}
\end{figure}

\begin{figure}
 \scalebox{0.35}{\includegraphics{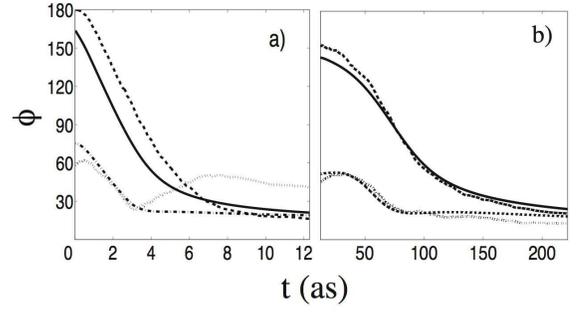}}
 \caption{\label{fignew:H} 
  Averaged angles as a function of time over
 the $\alpha=\1E$ ensemble. Solid: angle $\phi$ between $\vec
 F_{21}$ and $\vec p_{1}$ in a) and between $\vec
 F_{31}$ and $\vec p_{1}$ in b); Dashed: angle $\phi$ between $\vec
 F_{21}$ and $\vec p_{x,1}$ in a) and between $\vec
 F_{31}$ and $\vec p_{x,1}$ in b); Dash-dotted: angle $\phi$ between $\vec
 F_{12}$ and $\vec p_{2}$ in a) and between $\vec
 F_{13}$ and $\vec p_{3}$ in b); Dotted: angle $\phi$ between $\vec
 F_{12}$ and $\vec p_{x,2}$ in a) and between $\vec
 F_{13}$ and $\vec p_{x,3}$ in b).}
 \end{figure}

 Describing the collisions using the rate of change of the single
 electron Hamiltonians has the advantage that the effect of the
 nucleus is ``folded in''.  As a result both, the early collision which
 takes place close to the nucleus and the latter one which takes place
 far away from it (for both ensembles $\1E$ and $\2E$), exhibit exactly
 the same pattern, see
 \fig{wignerhammean}.  This becomes even clearer when one compares
 \fig{wignerhammean} with the probability densities of the momentum
 component along the $x$-direction for all three electrons in
 Figs.~\ref{fig7:px1}, \ref{fig9:px2} and \ref{fig11:pxy3}.  The
 momentum along the $x$-direction of the impacting electron 1 in
 ensemble $\1E$ switches form decreasing to increasing at times 3.5~as
 for the 12 collision and 79~as for the 13 collision.  This is a
 consequence of $\dot {H}_{1}$ switching sign at 2.4~as for the 12 and
 at 79~as for the 13 collision.  The time of 3.5~as, where $p_{x,1}$
 starts to increase in the 12 collision, is different from the time
 2.4~as, where $\dot {H}_{1}$ switches sign, while both times are the
 same in the 13 collision.  The reason is that for the 12 collision
 the $x,y,z$-coordinates are not equivalent, with the transfer of
 momentum taking place mainly along the $x$-direction, while for the
 later 13 collision the $x,y,z$-coordinates are almost equivalent.
 This is illustrated in \fig{fignew:H}, with $\vec{p}_{1} \cdot
 \vec{F}_{21}=0$ at 2.4~as and $\vec{p}_{x,1} \cdot \vec{F}_{21}=0$ at
 3.5~as.  The nucleus has a significant effect on the 12 collision
 while it has a small one on the 13 collision as seen by the more
 prominent increase of $p_{x,1}$ at 79~as when compared to its
 increase at 3.5~as.  The change with time of $p_{x,1}$ during the 12
 collision is due to $\vec{F}_{21}$ {\it and} the $-\partial
 {H_{1}}/\partial \vec{r}_{1}$ force from the nucleus, while in the 13
 collision the change of $p_{x,1}$ is mainly due to $\vec{F}_{31}$.
 In Figs.~\ref{fig9:px2}  and \ref{fig11:pxy3} we see that $p_{x,2}$ and
 $p_{x,3}$ increase during the time the respective electrons 2 and 3
 are impacted by electron 1, in agreement with $\dot{ H}_{2}>0$ and
 $\dot{ H}_{3}>0$ during the 12 and 13 collisions.  Similar
 conclusions can be drawn for the $\alpha=\2E$ ensemble.


\subsection{Inter-electronic angles}

\begin{figure}
\scalebox{0.35}{\includegraphics{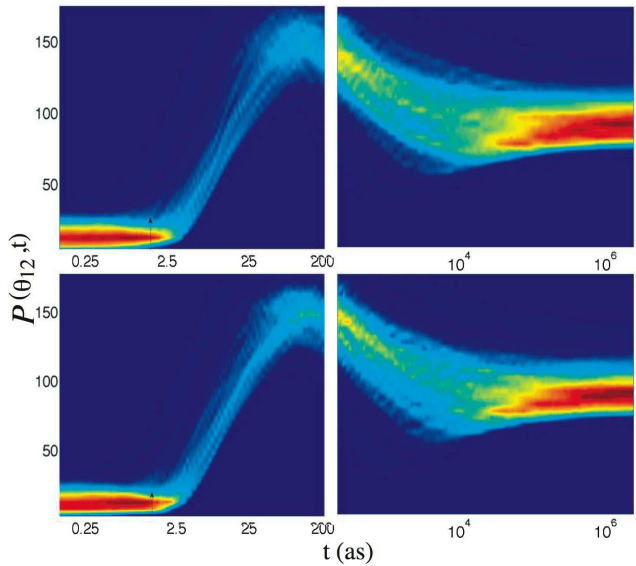}}
\caption{\label{fig4:theta12} Same as Fig.~\ref{fig7:px1} but for the
probability density of the inter-electronic angle $\theta_{12}$,
$\mathscr{P}_{\alpha}(\theta_{12},t)$.  The arrows indicate the time
of the collision $t_{ij}$.}
\end{figure}

 Finally, we discuss the time evolution of the interelectronic angles.
 For large times and quasi-free motion $\mathbf r_{i} \propto \mathbf
 p_{i}t$, the inter-electronic angles refer to the relation between
 positions as well as momenta of the electrons.  The dynamics in the
 angle is governed by two principles:
\begin{itemize}
 \item[(A)]
    Collisions between two electrons lead to a minimum of the angle
    $\theta_{ij}$ between the participating electrons $i$ and $j$,
    i.e., $\theta_{ij}(t_{ij})\approx 0$, if the collision happens at
    time $t_{ij}$.
 \item[(B)]
    Electrons tend to move away from each other minimizing their
    mutual repulsive interaction.  This leads to an interelectronic
    angle of $180^{\circ}$, if none of the electrons suffers a
    collision through the third electron.
\end{itemize}
 With these two principles, we recognize in \fig{fig4:theta12} the
 first collision early on (small angle, criterion A) and we also infer
 that a second collision happens around 65\,as, but \textit{not}
 between electrons 1 and 2, since suddenly the increase of their
 mutual angle towards $180^{\circ}$ (criterion B) is stopped and
 $\theta_{12}$ shrinks again towards its final value of $90^{\circ}$,
 giving rise to the ``T-shape'' structure of the three escaping
 electrons \cite{ER2}.  This is true for both collision sequences,
 (12,13) and (12,23).  The first collision happens in both cases
 between electrons 1 and 2, and then electron 3 imposes a second
 collision with one of the partners forming the angle in
 \fig{fig4:theta12}, namely with electron 1 (upper panels) and
 electron 2 (lower panels).  Finally, since in both cases the last
 colliding electron pair is not the 12, $\theta_{12}$ approaches
 $90^{\circ}$.
 
\begin{figure}
\scalebox{0.35}{\includegraphics{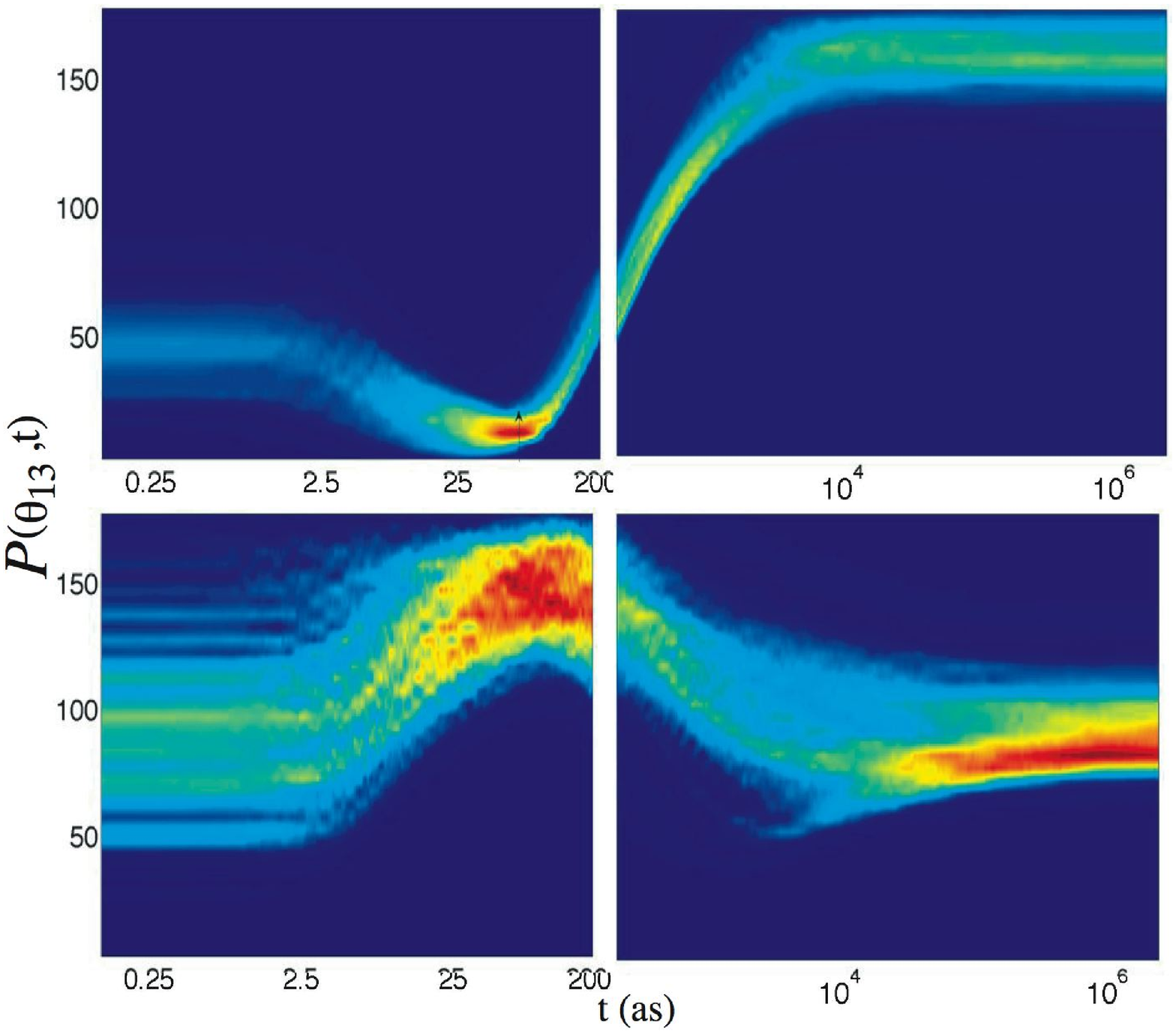}}
\caption{\label{fig5:theta13}Same as in \fig{fig4:theta12} but for
the inter-electronic angle $\theta_{13}$.}
\end{figure}

\begin{figure}
\scalebox{0.35}{\includegraphics{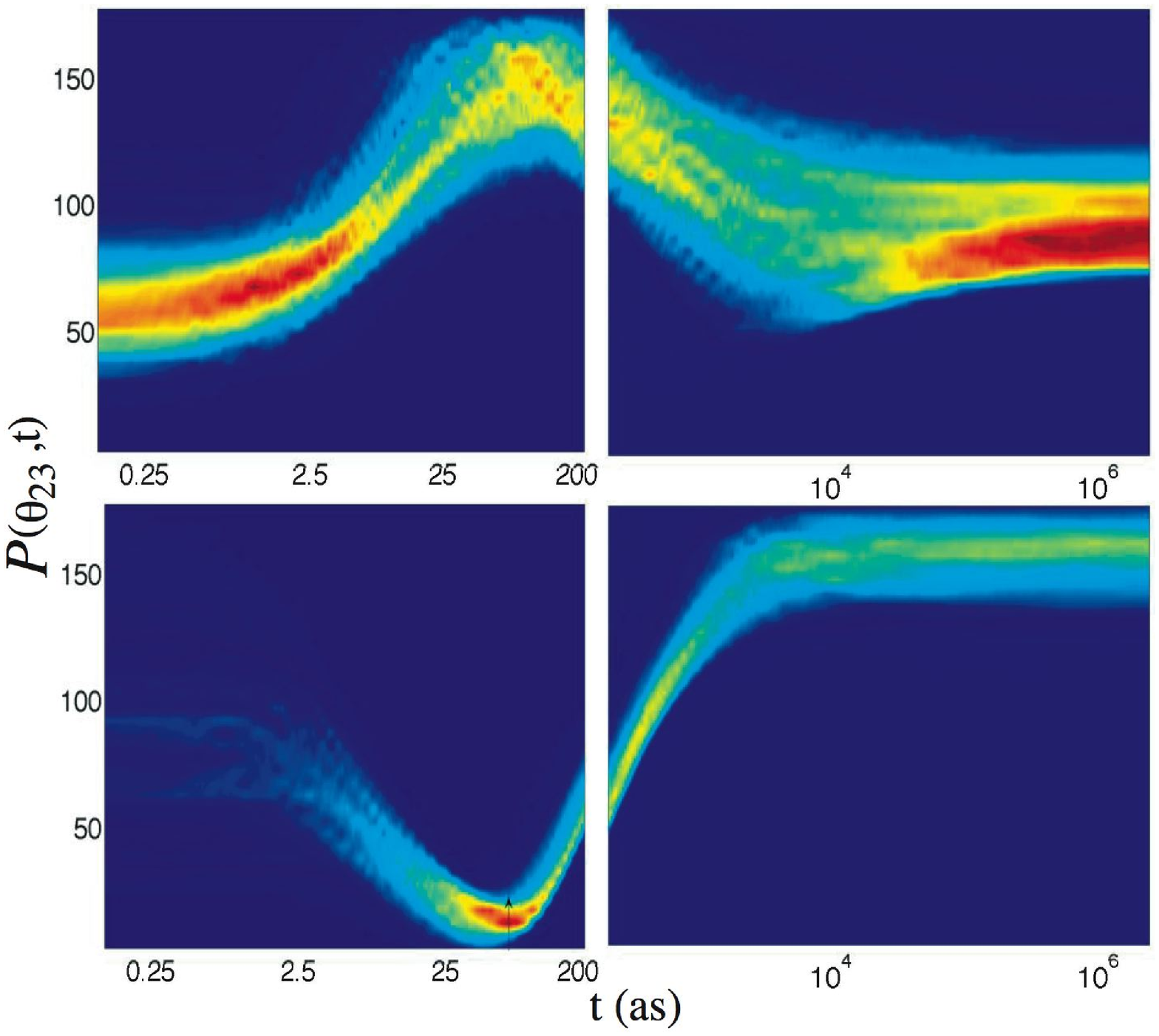}}
\caption{\label{fig6:theta23}Same as in \fig{fig4:theta12} but
for the inter-electronic angle $\theta_{23}$.}
\end{figure}

 The evolution of $\mathscr{P}(\theta_{13})$, \fig{fig5:theta13}, and
 $\mathscr{P}(\theta_{23})$, \fig{fig6:theta23}, differs much more for
 the respective two sequences (upper and lower panels).  However,
 there is a similarity across the two figures, namely the pattern in the
 upper(lower) panel of \fig{fig5:theta13} is similar to that in the
 lower(upper) panel of \fig{fig6:theta23}.

 The reason is that in the case of $\theta_{13}$ only the $\alpha=\1E$ 
 ensemble (upper panel of \fig{fig5:theta13}) leaves the
 visible imprint of a collision, bringing $\theta_{13}$ close to zero
 while it rapidly approaches $180^{\circ}$ afterwards, since it is the
 last collision and electrons 1 and 3 move away from each other
 afterwards.  The same, but now for electrons 2 and 3, is true for the
$\alpha=\2E$ ensemble (lower panel of \fig{fig6:theta23}).
 
 In the other two panels (lower panel of \fig{fig5:theta13} and upper
 panel of \fig{fig6:theta23}) one recognizes with the sudden turn away
 from $180^{\circ}$ for the respective interelectronic angle a
 collision with the third electron (criterion B), in case of
 \fig{fig5:theta13} the collision partner is electron 2 and in case of
 \fig{fig6:theta23} electron 1.
 
 Finally, it is worthwhile to note and understand the great variation
 in the width of the initial distribution for the angles across the
 three figures.  $\mathscr{P}(\theta_{12},0)$ is most strongly
 confined to values around zero since the 12-collision happens at an
 early time and electron 1, having absorbed the photon energy, starts
 with relatively high velocity close to the origin (position of the
 nucleus).  In the short time (1.7~as) before the collision the
 momentum vector of electron 1 cannot change substantially, so
 electron 1 keeps its direction.
 
 The condition for the first collision, $\mathrm r_{1}(t_{12}) \approx
 r_{2}(t_{12})$, implies also that the angles $\theta_{13}$ and
 $\theta_{23}$ should be similar at early times.  This is indeed the
 case, comparing the upper left panels of Figs.~\ref{fig5:theta13}
 and \ref{fig6:theta23}, where in both cases the maximum of the
 initial distribution is around $50^{\circ}$, while for the lower left
 panels the widely spread initial distribution is centered about
 $90^{\circ}$.  The latter indicates no clear preference in the
 initial mutual angle between electrons 2 and 3 reflecting the
 expectation value of an uncorrelated (product) wavefunction for the
 Lithium ground state as used here.  Moreover, one should keep in mind
 that whenever electron 3 is involved one would expect a wider
 distribution due to the larger size of the $2s$ initial electron
 density compared to the $1s$ density for electrons 1 and 2.
 
 \section{Conclusions}
  We have investigated the two
  main collision paths the three electrons follow to escape to the 
  continuum  from the
  ground state of Li after single photon absorption on an ensemble level.  Studying the classical
  probability densities for the two ensembles of trajectories
  corresponding to the two main collision sequences we were able to
  identify the traces these sequences leave on the classical
  probability densities.  Furthermore, we could show that each
  of the two ensembles has unique manifestations on
  the ensemble average level which clearly distinguish one from the
  other.  Being able to distinguish the two main attosecond time scale
  collision sequences on an ensemble level holds promise for a future
  \textit{direct} observation of these collision sequences with the
  advancement of ultrashort laser technology.


\end{document}